# Learning Continuous User Representations through Hybrid Filtering with doc2vec

Simon Stiebellehner[*]
Vienna University of Economics and Business
University College London
simon.stiebellehner.16@ucl.ac.uk

Jun Wang
University College London
MediaGamma Ltd
j.wang@ucl.ac.uk

Shuai Yuan
MediaGamma Ltd
shuai.yuan@mediagamma.com

## ABSTRACT

Players in the online ad ecosystem are struggling to acquire the user data required for precise targeting. Audience look-alike modeling has the potential to alleviate this issue, but models' performance strongly depends on quantity and quality of available data. In order to maximize the predictive performance of our look-alike modeling algorithms, we propose two novel hybrid filtering techniques that utilize the recent neural probabilistic language model algorithm doc2vec. We apply these methods to data from a large mobile ad exchange and additional app metadata acquired from the Apple App store and Google Play store. First, we model mobile app users through their app usage histories and app descriptions (user2vec). Second, we introduce context awareness to that model by incorporating additional user and app-related metadata in model training (context2vec). Our findings are threefold: (1) the quality of recommendations provided by user2vec is notably higher than current state-of-the-art techniques. (2) User representations generated through hybrid filtering using doc2vec prove to be highly valuable features in supervised machine learning models for look-alike modeling. This represents the first application of hybrid filtering user models using neural probabilistic language models, specifically doc2vec, in look-alike modeling. (3) Incorporating context metadata in the doc2vec model training process to introduce context awareness has positive effects on performance and is superior to directly including the data as features in the downstream supervised models.

## 1 INTRODUCTION

10 years ago, the launch of the first iPhone heralded the era of the smartphone. Nowadays, more than 80% of British adults own one and collectively look at their screens more than a billion times a day [20]. Over the last years, the steep climb in smartphone usage has mainly been driven by mobile apps [8]. Naturally, this has made mobile applications a popular advertising medium, resulting in ever larger parts of advertising budgets being allocated to advertisements in apps (in-app ads) [10]. Online advertising in general and mobile/in-app advertising in particular allow for unprecedented targeting of users, such as geospatial targeting, which promises to increase ad effectiveness [14].

In-app ad space can be (1) directly sold to advertisers, (2) dynamically allocated through the publisher's advertising network (ad network), (3) or through a combination of both. Ad networks are organisations that aim to establish a connection between publishers and advertisers. They typically act as cross-functional entities that aggregate publishers' inventory and sell it in bundled form to advertisers to match their demand. Matching of supply and demand is facilitated by technology platforms that are referred to as "supply-side" and "demand-side" platforms. While supply-side platforms (SSP) enable automation of selling of ad inventory (publisher-side), demand-side platforms (DSP) are their counterparts on the advertiser-side that automatize purchasing of ad space. Typically, SSPs and DSPs are connected through intermediaries acting as brokers ultimately performing the matching of supply and demand: ad exchanges (ADX). Generally, ad exchanges take the role of aggregators that facilitate purchasing and selling of ad space across advertising networks. For reasons of efficiency, online advertising transactions are increasingly being automated through the usage of ADX. On ad exchanges, selling and buying of ad space happens on a per-impression basis, in real-time and through an auction process. Therefore, this process is commonly referred to as "real-time bidding" (RTB) [5].

A large and strongly growing part of in-app ads are allocated through RTB [6]. In real-time bidding, single ad impressions are usually sold to the highest bidding advertiser in fractions of a second through auctions. Naturally, prerequisite for a successful matching of publishers' and advertisers' needs is the provision of data that specifies both supply and demand. Consequently, when a user generates an impression in an app, the corresponding publisher sends an ad request containing user and context-related information to the ad exchange. The ad exchange then sends a bid request to advertisers, requesting the bid and an advertisement that is displayed in case the auction is won. When the ad exchange finds a match based on both sides' provided data, the auction winner's advertisement is displayed and he pays the corresponding price.[1] This price may strongly depend on the characteristics of the user that generated the impression. Advertisers are willing to pay a premium for accurate targeting of users as it enables them to minimize wastage. Therefore, the publisher side puts great effort into collecting high-quality user data, which is crucial for targeting. However, this is a challenging task as users are typically reluctant to provide data about themselves. Nevertheless, the fact that detailed information is available for some users enables the use of audience look-alike modeling techniques to conclude on characteristics of others. Thereby statistical and machine learning techniques are used to, at least temporarily, impute missing user data with predictions derived from other users. Approaches to look-alike modeling come from a wide range of methods [34][16][22]. However, what all approaches have in common is that, the more data there is available and the higher the quality of the data, the better these algorithms can recognize patterns and perform the predictive task.

---

[1]We refer to [43] and [39] for a comprehensive elaboration of this process.

The improvement of look-alike models through maximizing quantity and quality of user data is directly related to this research. We aim to optimize the representation of users through data enhancement in order to maximize the performance of existing supervised machine learning models for look-alike modeling that predict gender and age of users in an RTB context. We do so by taking a recommendation system stance to user modeling, arguing that users' behaviors reflect their characteristics. In the context of in-app advertising, user behavior is expressed through the usage of mobile applications, i.e. users' app usage histories. This data would suffice to compute user and app representations following a collaborative filtering approach. However, in order to obtain a more fine-grained picture of users we acquire metadata of apps they used from the Google Play store and Apple App store. This supplementary app information would allow us to take a content-based filtering approach. Still, the weaknesses of both collaborative and content-based filtering are well-known and critical [1]. Therefore, we propose two novel hybrid filtering approaches that aim to overcome these issues and eventually improve the representation of users. We model users as (1) sequences of descriptions of apps they previously used (user2vec) and, additionally, (2) user and app metadata (context2vec). We do so by employing techniques for feature extraction from text, specifically the neural probabilistic language model algorithm doc2vec, to generate numeric vector representations of users in a hybrid filtering way. Our proposed methods overcome key shortcomings of existing approaches and close notable gaps in literature: (1) current techniques require crude forms of aggregation to model users, deteriorating the quality of user representations. (2) The application of neural probabilistic language models and especially doc2vec in hybrid filtering is yet widely unexplored. (3) The inclusion of contextual metadata in model training to introduce context awareness has not been investigated so far. (4) Hybrid filtering user models using neural probabilistic language models in look-alike modeling are yet entirely uncovered in literature. We answer three research questions (R1, R2, R3) in order to assess the performance of our proposed approaches both qualitatively in a recommendation system setting and quantitatively by using them as features in named machine learning models for look-alike modeling:

- **R1:** In a recommendation system setting, does modeling users as sets of descriptions of items they previously used (user2vec) and additional metadata (context2vec) improve the quality of recommendations provided by the system compared to only using non-descriptive user-item interactions?

- **R2:** Does modeling users as sets of descriptions of items they previously used and using the generated feature vectors in a downstream supervised task of user look-alike modeling improve predictive performance?

- **R3:** Can additional contextual user and item metadata be included in user models so that performance of downstream supervised models is improved further?

In the remainder of this paper, we first provide the reader with the technical preliminaries required for the understanding of the methods used in this research (section 2). Then, we discuss related work that has been done in the field and particularly focus on the application of neural probabilistic language models in recommendation systems (section 3). Subsequently, we describe the methodology of this work (section 4), including details on our approaches, their contributions, the experiments as well as the data. Finally, we present and discuss the results (section 5) and provide a conclusion (section 6).

## 2 PRELIMINARIES

### 2.1 Neural Probabilistic Language Models

A neural probabilistic language model (NPLM) is a type of language model that is based on a neural network, leveraging its capability to model high-dimensional discrete distributions [3]. The idea of NPLMs is based on the distributional hypothesis, suggesting that words appearing in a similar context are similar in meaning [35]. This is put into practice by using a neural network to extract linguistic similarities and semantic information from word co-occurrences, resulting in low-dimensional, fixed-length numeric vectors (neural text embeddings). There exists a variety of state-of-the-art NPLM algorithms with word2vec [26] and its extension doc2vec (paragraph2vec, sentence2vec) [19] arguably being the most prominent representatives. Both algorithms have shown to be of great value across a variety of domains, including machine translation [25] and sentiment analysis [37][19]. We provide a brief overview of word2vec and doc2vec and refer the interested reader to the original publications for details.

### 2.2 word2vec

The word2vec algorithm uses a shallow neural network to generate distributed, numeric vector representations of words in multi-dimensional space. Originally, two similar neural network architectures to generate these embeddings were proposed: continuous bag of words (CBOW) and skip-gram (SG). Since CBOW and SG follow the same principle but in reversed ways, we elaborate CBOW in more detail and limit the explanation of the skip-gram model to the points that differ notably from CBOW. First, each token of the vocabulary $V$ of the text corpus is mapped to a randomly initialized weight vector, which constitutes a column in the weight matrix $W$ of the neural network. A given word is denoted as $w_t$ while its context $c_{t,n}$ of window size $n$ is composed of $(w_{t-n}, ..., w_{t+n})$. In the CBOW model, a target word $w_t$ is predicted given its context $c_{t,n}$. The CBOW neural network architecture to achieve this is fully connected and has one input layer, one hidden layer and one output layer.[2] Assuming a single-word context for simplification (only one $W$ at the input layer instead of $n$ weight matrices in a multi-word context), the input layer $x$ is a one-hot encoded vector of $|V|$ dimensions. The hidden layer size is $N$ and can be treated as a parameter that determines the dimensionality of the word embeddings. Terming the weight matrix between input and hidden layer $W$ and the one between hidden and output layer $W'$, highlighting

---

[2] For skip-gram, the architecture would be exactly reversed, showing one input vector of dimension $|V|$ and $n$ output vectors of length $|V|$.

Learning Continuous User Representations through Hybrid Filtering with doc2vecthat there is a linear activation function at the hidden layer and a softmax function at the output layer, computations are as follows:[3]

$$b = W^T x \quad (1)$$

$$z = b^T W' \quad (2)$$

$$P(w_t|c_{t,n};\theta) = y_t = \frac{exp(z_t)}{\Sigma_{t'=1}^{V} exp(z_{t'})} \quad (3)$$

In the optimization process, the objective function, or equivalently its log-likelihood form, is maximized[4]:

$$argmax_\theta \prod_{(w_t, c_{t,n}) \in D} P(w_t|c_{t,n};\theta) \quad (4)$$

The optimization problem can be solved with, for instance, stochastic gradient descent (SGD) with back propagation. After training, the hidden states $b$ of each word represent the $N$-dimensional word embeddings.

The formulation of the skip-gram model is exactly reversed to CBOW. In SG, each context word is predicted from a target word. Therefore, while there is always only one input vector $x$ of length $|V|$, there are $n$ output vectors (context window size) of length $|V|$ where each represents a Multinomial distribution. The formulation of the objective function only differs marginally from CBOW by accounting for the reversed problem formulation:

$$P(c_{t,n}|w_t;\theta) = y_{t,c_{t,n}} = \frac{exp(z_{c_{t,n},t})}{\Sigma_{t'=1}^{V} exp(z_{t'})} \quad (5)$$

$$argmax_\theta \prod_{(w_t, c_{t,n}) \in D} P(c_{t,n}|w_t;\theta) \quad (6)$$

### 2.3 doc2vec

Essentially, doc2vec extends word2vec by another input to the model, which is a unique, randomly initialized numeric vector, representing the document the target word is part of. This allows for considering each document as a dedicated element in the learning process. Similar to word2vec, the authors provide two variants of doc2vec that realize this in slightly different ways: distributed memory (DM) and distributed bag of words (DBOW). In a DM model, the document vector $x_p$, where $p$ is the number of documents in the corpus, is treated as an additional input analogue to a word. Therefore, $b$, the hidden states, now depend on both word input vectors and the document input vector. Since a document vector is shared among all words/contexts sampled from that document and word vectors are shared across all documents these words occur in, this can be thought of as realizing a memory function. At the hidden layer, the inputs are concatenated or averaged, depending on the specific implementation. Finally, they are passed to the softmax function at the output layer.

The DBOW variant is simpler and trains faster. In this model, context words are ignored at input level, but only the document vector is used as input. Then the model is trained through predicting a word sampled from a randomly chosen window from the input document.

---

[3]Multi-word context: $b = \frac{1}{n} W^T(\Sigma_i^n x_i)$
[4]$D$ denotes the document corpus as a set of all word/context pairs.

## 3 RELATED WORK

Traditionally, user modeling is the domain of recommendation systems. The most prevalent techniques in this field are content-based filtering (CBF) and collaborative filtering (CF). Furthermore, these methods have been blended into hybrid approaches (HF) [7]. Content-based filtering is characterized by the utilization of item and user data to construct corresponding profiles. The system then recommends items that match users' profiles (*item-to-user*) [1]. In contrast, collaborative filtering is independent of content, but solely relies on the usage patterns of users and items. Therefore, recommendations are purely based on the similarity (e.g. cosine similarity) of user or item vectors (*user-to-user*, *item-to-item*) of a user-item matrix (history matrix) [40]. In the last years, techniques originating from natural language processing (NLP) have gained popularity in the field of recommendation systems. In content-based filtering, NLP methods are used to generate numeric representations from textual elements of user and item profiles. Less intuitively, NLP techniques also find application in collaborative filtering, such as for the purpose of reducing the dimensionality and sparsity of user and item vectors.

Initially, the bag of words (BOW) model found popular application in CBF [27][23][21]. In order to overcome this representations' lack of incorporation of term importance, the term weighting scheme TF-IDF[5] was used to encode textual corpora in content-based recommendation tasks [42][31]. Both BOW and TF-IDF typically result in high-dimensional, sparse vectors. Furthermore, both methods assume independence of terms and fail to capture complex elements of language such as semantic similarities. Latent semantic models such as the SVD-based[6] latent semantic analysis (LSA) [9] and the probabilistically motivated latent dirichlet allocation (LDA) [4] build upon BOW and TF-IDF and attempt to eliminate their weaknesses in text feature extraction. Furthermore, LSA and LDA have also found application in collaborative filtering [13][36][18][33][41]. However, critical weaknesses such as ignorance of (word) order, requirement of prior knowledge of the number of topics in the corpus, the implicit assumption of exchangeability of words and unfavorable scaling properties remain. These drawbacks have spurred the invention and development of neural probabilistic language models. NPLMs have shown performance competitive and often superior to state-of-the-art methods while being highly efficient [24][26][2]. The great popularity of neural text embeddings in NLP and the conceptual similarity of generating embeddings from sequences of words and generating embeddings from arbitrary sequences have motivated researching NPLMs' application in fields only distantly related to natural language processing such as computer vision [38] and recommendation systems [2][12][30][32][28][29].

There are numerous reasons why the application of NPLMs in recommendation settings may be attractive. First, NPLM algorithms generate dense, fixed-length vectors of arbitrary dimensions, solving the sparsity problem in collaborative filtering and extracting features from textual elements in content-based filtering. Second, due to the usage of shallow neural networks, they are fast to train even at large scale. Third, vectors generated by NPLMs are intuitively interpretable when projected onto low-dimensional space. Fourth, neural network language models consider the order of elements. Overall, NPLMs have shown quality of results en par or even superior to existing methods in recommendation settings [2][29] while being highly efficient. The application of NPLMs in recommendation systems may not be obvious at first. The idea is that the principle of the distributional hypothesis can be applied to arbitrary sequences whose elements are related to one another following a pattern. Since this is the case for usage and purchasing histories, neural network language models are able to generate sensible vector representations of users and items from that data. Recommendations can then be inferred from generated embeddings by computing the normalized cosine similarity of embedding vectors.

---

[5]Term Frequency - Inverse Document Frequency
[6]Single Value Decomposition

Mapping a recommendation task to a format suitable for NPLM algorithms can be achieved in various ways. In [29], Musto et al. use word2vec in a content-based recommendation setting. They associate each item with a corresponding article from Wikipedia, whose words are used as input to the word2vec algorithm. Vectors of words contained in an item description are then aggregated by averaging to generate a vector representation of each item. Users are modeled as the centroids of all vectors of items they purchased. Similarly, the same authors use (stacked) summations to aggregate word embeddings on item and user level in [28]. These approaches have a number of drawbacks. First, they employ word2vec, which forces them to perform crude aggregations of vectors on higher levels. Aggregating embeddings in this form deteriorates their quality by, for instance, losing encoded order of elements. Second, they use Wikipedia descriptions of items instead of text specifically tailored to these items. Due to their encyclopedic nature, Wikipedia articles are written to generalize well. However, generalistic item descriptions may lead to failure to differentiate between items of the same type (e.g. different kinds of software being associated with the description of "software"). This indicates that their approach is unlikely to be applicable in practice.

Grbovic et al. [12] follow a pure collaborative filtering approach using word2vec, solely relying on the user-item matrix. Again, due to the use of word2vec they must employ some form of aggregation to model users. They do so by considering users as centroids of vectors of items they used. Similar approaches are taken by Barkan et al. [2] and Ozsoy [30]. All these pure collaborative filtering approaches exclusively rely on the user-item matrix. This poses a problem when items have only been used a few times or not at all (item cold start) or when new users enter the system (user cold start). Furthermore, they do not allow for direct item-to-user recommendations.

Only very little notable research has yet explored the potential of applying NPLMs tailored to encode blocks of text in recommendation systems and user modeling. In [32], Phi et al. successfully use doc2vec in a collaborative filtering task. They do so by treating items as words that form a document, which represents a user. We refer to this approach as **d2v:CF**. By using doc2vec they overcome the issue of modeling users through crude aggregations, such as summation or averaging, which are required when using word2vec. Instead, both item and user vectors are directly generated by the doc2vec algorithm. The similarity of users depends on the items they previously used. Likewise, the similarity of items results from their co-occurrence in users' usage histories. The method of Phi et al. essentially represents the state of the art of using doc2vec in a recommendation system setting, however, still suffers from the typical weaknesses of collaborative filtering.

The presented applications of neural network language models in recommendation systems show that current approaches are flawed or come with significant drawbacks. Furthermore, it highlights current unexplored areas of research:

**(1)** Currently, word2vec is still the dominant algorithm among NPLMs for recommendation tasks. However, word2vec does not provide a natural way of generating vectors for blocks of text. This leads to the necessity of crude aggregations for modeling users and even items in some settings, deteriorating embedding quality.

**(2)** What is more, yet NPLMs have almost exclusively been applied in collaborative filtering. So far, applications in content-based filtering used suboptimal datasets (Musto et al. [29] using Wikipedia descriptions) and/or employed cumbersome and questionable stacked aggregations to generate item and user vectors. Especially NPLMs tailored to blocks of text (doc2vec) lack investigation in recommendation tasks other than collaborative filtering. To the best of our knowledge, applications of doc2vec in hybrid filtering are entirely unexplored so far.

**(3)** On top of item descriptions, often there is contextual metadata on items and users available. However, yet research lacks investigation of incorporating context metadata into training of NPLMs to introduce context awareness.

**(4)** Furthermore, user models generated through hybrid filtering methods employing NPLMs have not yet been used in look-alike modeling.

The doc2vec algorithm offers a natural way of directly generating embeddings for items and users without the need for aggregation. What is more, it allows for including basically arbitrary user and item information, facilitating the construction of user and item models following a hybrid filtering approach. A hybrid filtering technique using doc2vec enables joint encoding of usage patterns and item descriptions. The novel approaches we propose in this research aim to exploit these properties to eliminate the highlighted flaws of current techniques and contribute to closing the identified gaps in literature.

## 4 METHODOLOGY
### 4.1 Proposed Approaches

We propose two hybrid filtering approaches based on doc2vec that remedy the discussed shortcomings of existing methods and allow for investigation of the effect of joint encoding of heterogeneous information. The first approach, termed "user2vec", uses item descriptions and usage histories to model users while the second, "context2vec", additionally utilizes further metadata on items and users in an attempt to incorporate context into the model.

*4.1.1* **user2vec**. We showed that recommendation systems that are based on NPLMs yield state-of-the-art performance. However, existing approaches limit themselves either to content-based filtering or to collaborative filtering. Content-based filtering techniques so far employed word2vec, leading to the issue of deteriorating embedding quality through crude aggregations of vectors on user level. Also, their purely content-based nature does not enable them to leverage information on user-item interactions. On the other hand, existing collaborative filtering methods, such as d2v:CF by Phi et al. [32], fail to capture similarities of items that have not been used together. For instance, a car racing and a motorcycle racing game are similar by nature, however, CF may be unable to capture this in case of one of them being little popular hence only showing few usages. These deficits stress the necessity of formulating a hybrid filtering method based on doc2vec, combining CF and CBF. A thorough combination would provide a more reasonable measure of similarity by taking into account both similarity of content and similarity of usage patterns. Therefore, we propose user2vec. This approach combines user-item usage information and item descriptions by jointly encoding them using doc2vec. First, analogue to standard content-based filtering methods, we consider each item $j$ to be represented by its textual description $d_j$.

$$d_j = [this, text, describes, item_j] \qquad (7)$$

Then, in order to jointly encode user-item interactions and item descriptions, we model each user $u_k$ as sequence of descriptions of items they previously used $F$.

$$u_k = \{d_j, ..., d_n\} \; for \; j \in F \qquad (8)$$

We can then use this corpus of models $u_k$ of length $|U|$ (number of unique users) to train a doc2vec model as described in the preliminaries section by considering each user a document. Subsequently, the trained model allows us to perform inference to obtain vectors for arbitrary users and items. Through this novel formulation, the generated embeddings express the entire pool of heterogeneous information that is typically available in recommendation systems (user-item matrix, item descriptions) in a clean and coherent way. The user2vec approach has numerous advantages, such as (1) establishing and encoding the relationship between descriptive item information and usage patterns, (2) avoiding suboptimal forms of vector aggregations, (3) providing user-to-user, item-to-item and item-to-user recommendations through users and items sharing the same vector space and (4) alleviating the item cold start problem.

*4.1.2 context2vec.* Apart from item descriptions and usage histories, often there are further variables available that provide context information in the form of meta data. This yields the potential of ultimately providing a more holistic picture by encoding the relationship between the entities in context. However, current approaches do not include additional context data in training of the embedding models. Therefore, we investigate establishing context awareness in hybrid filtering using doc2vec by incorporating supplementary metadata with the aim of improving the vector representations of users. Concretely, this is realized through *extending* user2vec. In this approach, we model items $i_j$ as sequences of their textual descriptions $d_j$ and metadata $m_j^i$.

$$i_j = \{d_j, m_j^i\} \qquad (9)$$

Subsequently, we formulate user models as sequences of descriptions and corresponding metadata of items $i_j$ they previously used. Moreover, we add meta information $m_k^u$ (e.g. city, operating system) of the respective user $u_k$.

$$u_k = \{i_j, ..., i_n, m_k^u\} \; for \; j \in F \qquad (10)$$

Again, after training a doc2vec model on the corpus consisting of all user representations $u_k$, we can perform inference to compute any user and item embedding vector. On top of establishing the relationship between usage patterns and item descriptions, this novel approach attempts to introduce context awareness through incorporating user and item meta information. The joint encoding of all available data promises to provide a complete picture of users in the form of fixed length numeric vectors. Furthermore, context2vec does not only alleviate the item cold start problem (as in user2vec), but also eases the user cold start by building a basic user profile through user metadata without requiring any information on item usage.

### 4.2 Experiments

In our experiments, we attempt to answer three related research questions (R1, R2, R3) with the purpose of assessing the contributions of our proposed approaches in recommendation system settings and in predictive look-alike modeling tasks.

*4.2.1 **R1**: In a recommendation system setting, does modeling users as sets of descriptions of items they previously used (user2vec) and additional metadata (context2vec) improve the quality of recommendations provided by the system compared to only using non-descriptive user-item interactions (d2v:CF)?* In this research, we model users by taking a recommendation system stance. Therefore, it is only natural that we perform qualitative assessment by comparing the quality of recommendations provided by our proposed approaches for different tasks: item-to-item, user-to-user and item-to-user recommendations. Furthermore, we benchmark them against the current state of the art of applying doc2vec in recommendation systems, Phi et al.'s d2v:CF [32].

*4.2.2 **R2**: Does modeling users as sets of descriptions of items they previously used and using the generated feature vectors in a downstream supervised task of user look-alike modeling improve predictive performance?* First, we generate feature vectors from item descriptions using a number of baseline approaches (TF-IDF, LSA, LDA, word2vec[7]) and aggregate them on user level by computing the centroid of vectors of items a user previously used. Additionally, we include the collaborative filtering method d2v:CF and the novel hybrid filtering approach user2vec in our evaluation. Finally, the generated vector representations are used as additional features in supervised machine learning models for look-alike modeling. These models and the evaluation of our approaches are discussed in subsequent sections.

*4.2.3 **R3**: Can additional contextual user and item metadata be included in user models so that performance of downstream supervised models is improved further?* It is investigated whether using the context-aware user model (context2vec) in the downstream supervised task further improves predictive performance. We compare the results to the best performing models from R2. Furthermore, we concatenate the feature vectors generated by these context-unaware models with conventional categorical and continuous representations of the additional metadata used in context2vec.

---

[7]Analogue to [29], due to word2vec generating word vectors, we aggregate them on user level by computing the centroid of word vectors of words contained in all descriptions of apps a user previously used.

Thereby we directly include them as features in the supervised task without incorporating them in the doc2vec model training. This supplementary comparison allows us to additionally determine if including metadata in doc2vec model training performs better than including it as categorical or continuous numeric features directly in the supervised models in a conventional manner. Evaluation is conducted analogue to R2.

*4.2.4 Evaluation.* For R1, similar to [2], we first qualitatively investigate the performance in user-to-user, item-to-item and item-to-user recommendation tasks by drawing random samples from the user and item pool for each approach. Then the quality of recommendations is compared between methods. We assume high quality recommendations to be closely related to the sampled entities topic-wise. Additionally, in the course of evaluating the results of R2 and R3, we assess the contributions of these approaches and baselines to improvements in predictive performance of supervised machine learning models for look-alike modeling by including the generated user vectors as features. These look-alike modeling algorithms are developed and used in practice by the London-based ad-tech company MediaGamma. Their models predict gender and age group of users and provide a means to evaluate the vectors resulting from the baselines and our proposed approaches in practice. Since they frame the predictive task as a binary classification problem, they use 9 different logistic regression models, one for each gender and age group, predicting true/false. The labeled data (seed users) is split into train and test data. Then, each model's optimum parameters are chosen using grid search and k-fold cross validation on the train dataset. Eventually, the trained models' performance is evaluated on the test set. Since each of the 9 predictive tasks is a binary classification problem with considerably imbalanced and equally weighted classes, we choose AUC-ROC[8] as evaluation metric [15].

*4.2.5 Hyperparameter Optimization.* Since the user vectors are ultimately used in the downstream task of look-alike modeling, the hyperparameters of the algorithms used to generate them are optimized such that the average of the AUC-ROC metric across all predictive models is maximized. Due to the vast space of potentially optimum values, the great number of different models, the large size of the data and the consequential extensive training times, we limit the extent of the tuning procedure to knowingly highly important hyperparameters. For others we select default/medium-range values. For each hyperparameter that is subject to tuning, we define a reasonable range of values. Then we use grid search and k-fold cross validation to find the optimum values in the defined spaces.

*4.2.6 Data.* We use two different datasets. Our base dataset (RTB dataset) is provided by a large European mobile ad exchange and comprises of bid requests on ad impressions. The supplementary dataset containing metadata of mobile applications mentioned in the bid requests is acquired from the Google Play store and the Apple App store (app metadata).

The RTB dataset is a collection of bid requests on ad impressions in mobile applications, geographically limited to the United Kingdom. MediaGamma uses this dataset in their predictive machine learning models for look-alike modeling. However, they typically only consider the data of the preceding week for the weekly model training. This selection has shown to yield close-to-optimum results while reducing data size to a better manageable amount of a few terabytes (TB). We follow this approach and only use a single week in our experiments (29$^{\text{th}}$ May to 4$^{\text{th}}$ June 2017) to remain close to its practical purpose and at the same time reduce computational efforts to a reasonable extent. The data of one week contains approximately 2,000,000,000 samples. One sample is one bid request for ad space of the inventory of a mobile publisher recorded by the ad exchange. A part of the variables included in a bid request contains information on the user triggering the request, such as city and operating system (OS). However, there are two kinds of users: known users (seed users), which is the smaller part, and unknown users. Bid requests on ad impressions generated by seed users additionally contain data on gender and age of the user. Each bid request represents an interaction of a user, uniquely identified by the variable IFA, and a mobile application (item), identified by its bundleID. In this research, we are interested in modeling a user based on the apps he used. Therefore, we only consider unique interactions of users and apps, i.e. unique combinations of IFA and bundleID, leaving us with roughly 4,000,000 entries. The fact that global unique identifiers of apps, their bundleIDs, are contained in the RTB dataset provides us with the opportunity to obtain additional metadata on the corresponding mobile apps.

Therefore, mobile application metadata is acquired through official and inofficial APIs of the Google Play store and Apple App store using apps' bundleIDs. The resulting dataset comprises of the variables app description, genre, average rating, number of ratings, price and bundleID. We remove mobile apps whose description is in a language other than English (~5%) since dealing with textual corpora of different languages in machine learning introduces significant noise and is an active area of research by itself. Subsequently, the RTB dataset is joined with the metadata on bundleID, dropping all bid requests that are not related to a bundleID that could be scraped. Furthermore, we exclude users that have only used very few different apps (less than 3) as this lack of information would not allow us to conclude on their characteristics in a meaningful way. This leaves us with a dataset of approximately 2,150,000 unique interactions of users and apps as a combination of 555,000 unique users (IFAs) and 8,200 unique apps (bundleIDs).

We only make use of a limited number of variables available in the data. Whereas we use all scraped variables from the app metadata, we select only IFA, city and OS from the RTB dataset. This small selection is based on the fact that, since we aim to model users, only directly user-related data is relevant. Furthermore, among the user-related variables, some only contain one value (e.g. country, language), others mostly contain unique values (e.g. IP address) and some are highly unlikely to be relevant to differentiate users (e.g. OS version number) or already included in another form (e.g. region). For seed users there are age and gender information additionally available. However, since we intend to predict these variables in parts of our evaluation, we do not include them in the data used for model training. This leaves us with the following selection: bundleID, app description, genre, average rating, number of ratings, price, IFA, operating system, city. Common preprocessing

---

[8]Area Under the Curve - Receiver Operating Characteristic

Learning Continuous User Representations through Hybrid Filtering with doc2vec

| App 1: | sports news app (com.scores365) | | |
|---|---|---|---|
| | **d2v:CF** | **user2vec** | **context2vec** |
| **SIM 1** | social game | sports news app | football news app |
| **SIM 2** | football news app | football news app | football news app |
| **SIM 3** | picture encryption app | sports news app | football news app |

Table 1: Apps most similar to App 1.

| App 2: | flashlight app (com.jiubang.fastestflashlight) | | |
|---|---|---|---|
| | **d2v:CF** | **user2vec** | **context2vec** |
| **SIM 1** | camera app | flashlight app | music streaming app |
| **SIM 2** | weather app | flashlight app | countdown app |
| **SIM 3** | foto editor | flashlight app | camera app |

Table 2: Apps most similar to App 2.

techniques are applied to this data: app descriptions are first tokenized, then stop words, short tokens and numbers are removed before each token is lowercased and lemmatized. Depending on the specific task, other variables are either binned and converted to strings or directly used as numeric or categorical features.

## 5 RESULTS & DISCUSSION
### 5.1 R1: Recommendation Setting

*5.1.1 **Item-to-Item***. Tables 1 and 2 show the top three recommendations[9] (i.e. most similar apps) for two randomly sampled apps in order of descending similarity. In both test cases, d2v:CF recommends mostly unrelated mobile apps. In contrast, user2vec suggests apps that are highly related to the benchmark app. The context2vec approach performs well for App 1, however, recommendations for App 2 seem to be unrelated. Overall, the incorporation of app descriptions into model training seems to put user2vec and context2vec at a notable advantage in this task. However, including metadata in addition to app descriptions, as in context2vec, may introduce noise that lowers the ability of the system to predict topic-related items again.

*5.1.2 **User-to-User***. Analogous to the item-to-item recommendation task, we randomly choose two users (User 1 - IFA: 2CD2C633-2FD5-43EF-AC27-69D4095AEB93, User 2 - IFA: 6161921F-BA2D-43DE-B07B-9A0F8E89154D) and compare derived recommendations. Tables 3 and 4 show the app usage histories of the users most similar to the benchmark users by normalized cosine similarity ("neighbors": SIM 1, SIM 2, SIM 3). These mobile applications can be considered recommendations to the benchmark users. The app usage histories of the neighbors partly overlap and are generally highly similar to benchmark users'. The approach d2v:CF seems to produce homogeneous recommendations, i.e. recommendations that are highly related to benchmark users' used apps. The user2vec method provides slightly more heterogeneous app suggestions such as a foto editing app or a music making game (User 1) that still seem to fit to the user profile. The recommendations of context2vec appear to be drifting further off. For example, it suggests a picture quiz game and a skill game to User 1, whose profile shows a strong interest in fashion and babies. *Moderate* heterogeneity in recommendations is not necessarily a disadvantage. While it may deteriorate accuracy [11], it contributes serendipity, whose absence is often criticized in systems that issue overly homogeneous suggestions.

[9]We replaced the original app names by descriptive names for better understanding.

| User 1: | fashion game, baby simulation game, styling game | | |
|---|---|---|---|
| | **d2v:CF** | **user2vec** | **context2vec** |
| **SIM 1** | fashion game, bakery game, fashion/makeup game | music making game, baby kitten game, baby simulation game, fashion game | skill game, fashion game, baby simulation game |
| **SIM 2** | baby kitten game, fashion game, baby simulation game, eating game | music making game, foto editor, jump & run game, baby simulation game | music streaming app, picture quiz game, fashion game, skill game, baby simulation game |
| **SIM 3** | fashion game, styling game, pet game, baby simulation game | music making game, screensaver app, fashion game, baby simulation game | music making game, screensaver app, fashion game, baby simulation game |

Table 3: App usage histories of users most similar to User 1.

| User 2: | card game, video chat, wallpaper app | | |
|---|---|---|---|
| | **d2v:CF** | **user2vec** | **context2vec** |
| **SIM 1** | card game, weather app, video chat, wallpaper app | video chat, wallpaper app, screensaver app | music streaming, video chat, wallpaper app |
| **SIM 2** | card game, wallpaper app, screensaver app | video chat, wallpaper app, screensaver app | card game, ebook app, wallpaper app, screensaver app |
| **SIM 3** | video chat, wallpaper app, wallpaper app | music streaming, video chat, wallpaper app | video social network, video chat, wallpaper app |

Table 4: App usage histories of users most similar to User 2.

| User 1: | fashion game, baby simulation game, styling game | | |
|---|---|---|---|
| | **d2v:CF** | **user2vec** | **context2vec** |
| **SIM 1** | decoration game | baby simulation game | card game |
| **SIM 2** | sound recorder | baby simulation game | baby simulation game |
| **SIM 3** | wallpaper app | baby simulation game | baby simulation game |

Table 5: Apps most similar to User 1.

*5.1.3 **Item-to-User***. For the qualitative assessment of item-to-user recommendations we use the same randomly sampled benchmark users as in the previous task to gain an additional element to compare the results to. Tables 5 and 6 show the recommendations. The results show that d2v:CF does not provide sensible recommendations whereas the recommendations of user2vec and context2vec seem reasonable. This result was expected as d2v:CF employs pure collaborative filtering, which does not allow for direct, reasonable assessment of similarity between users and items. In contrast, user2vec explicitly builds user and item profiles in the same vector space using app descriptions and, additionally, metadata in context2vec. This hybrid filtering nature puts them at a significant advantage by enabling direct item-to-user recommendations.

| User 2: | card game, video chat, wallpaper app | | |
|---|---|---|---|
| | d2v:CF | user2vec | context2vec |
| SIM 1 | styling game | wallpaper app | wallpaper app |
| SIM 2 | styling game | skill game | word game |
| SIM 3 | pet game | card game | image entertainment app |

Table 6: Apps most similar to User 2.

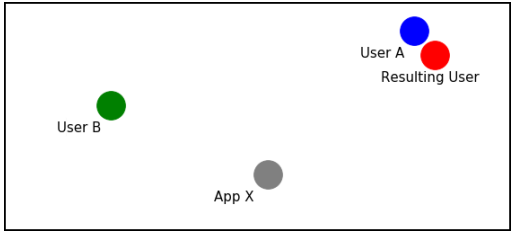

Figure 1: Projection of elements of arithmetic operations on user and app vectors generated by user2vec using t-Distributed Stochastic Neighbor Embedding (t-SNE).

Building user and item profiles and having them share the same vector space also allows for performing intuitive arithmetic operations on users and apps. For instance, in Figure 1 we exemplary perform the following simple mathematical operation using vectors[10] generated from user2vec:

$$User\ A = \{App\ X,\ App\ Y,\ App\ Z\}$$
$$User\ B = \{App\ Y,\ App\ Z\}$$
$$User\ A \approx User\ B + App\ X$$

Naturally, as our assessment is of qualitative nature and the size of the random samples is small, these observations cannot be taken as certain. Nevertheless, the results of this experiment indicate that, whereas the collaborative filtering approach d2v:CF yields good user-to-user recommendations, it performs poorly in the item-to-item and item-to-user case. In contrast to d2v:CF's pure reliance on user-item interactions, our novel hybrid filtering approaches, user2vec and context2vec, make use of app usage histories and app descriptions respectively additionally user and app metadata. As we have seen, this enables them to perform well in any kind of recommendation task, acknowledging R1. Furthermore, the possibility of performing arithmetic operations on users and apps fosters tangibility of the results and may find useful application in practice. Besides that, it is notable that the inclusion of metadata as in context2vec does not seem to give it an edge over only using app descriptions as in user2vec. This may be due to the additional metadata being of low quality, inadequate for this task or introducing noise.

## 5.2 R2: Hybrid Filtering in Look-Alike Modeling

Table 7 shows the absolute performance changes in AUC-ROC across all models and methods. It is notable that, overall, every method results in performance improvements compared to the zero-baseline (None). Averaging TF-IDF values across used apps for aggregation on user level outperforms all other approaches. This method relatively improves the zero-baseline by 1.91% AUC-ROC (1.08% absolutely) on average across all models. Therefore, R2 can be acknowledged: modeling users using descriptions of apps they previously used is beneficial for the predictive performance of the downstream supervised models.

---
[10]User A - IFA: 1DE8303E-FEB5-4BB3-8930-A4D18709667F | User B - IFA:b6645904-ce54-4c37-b3c0-702f2416a9cd | App X - bundleID: 909351158

| AUC-ROC | None | TF-IDF | LSA | LDA | word2vec | d2v:CF | user2vec |
|---|---|---|---|---|---|---|---|
| male | 58.55% | 0.95% | 0.68% | 0.32% | 0.87% | **1.31%** | 1.11% |
| female | 57.76% | 0.66% | 0.28% | 0.44% | 0.50% | **0.92%** | 0.58% |
| 18-24 | 61.12% | **0.69%** | 0.34% | 0.11% | 0.15% | 0.43% | 0.01% |
| 25-34 | 55.63% | **1.33%** | 0.68% | 0.55% | 0.73% | 1.00% | 0.97% |
| 35-44 | 52.78% | 0.73% | 0.44% | 0.04% | 0.54% | 0.45% | **0.79%** |
| 45-54 | 52.44% | **0.91%** | 0.36% | 0.04% | 0.52% | 0.24% | 0.29% |
| 55+ | 57.25% | 1.79% | 0.82% | 0.49% | 0.97% | **1.85%** | 1.30% |
| 18-34 | 56.54% | **1.29%** | 0.66% | 0.41% | 0.54% | 1.17% | 0.67% |
| 35+ | 56.03% | **1.38%** | 0.51% | 0.54% | 0.45% | 1.13% | 0.73% |
| Average | | 1.08% | 0.53% | 0.33% | 0.59% | 0.94% | 0.72% |

Table 7: R2. Absolute change compared to the zero-baseline (None) in AUC-ROC score for each task and method.

The benchmark approach d2v:CF shows results almost en par with TF-IDF. This is especially interesting as TF-IDF and d2v:CF model distinct aspects of the data. Whereas TF-IDF summarizes the content of app descriptions in a crude way, d2v:CF explicitly does not consider any meaningful textual content but only user-item interactions. In contrast, user2vec, which includes app descriptions in the modeling process, leads to smaller improvements than TF-IDF and d2v:CF. Since user2vec considers both user-item interactions and app descriptions (hybrid filtering) it was originally expected to perform best as it was thought to have the potential to encode most information jointly. What strengthened this believe was that it provided excellent results in the recommendation tasks of R1. However, the attempt to include more information in the model might have lead to the introduction of more noise, leading to poorer performance. This notion is confirmed by the results of the user-to-user task in R1, where user2vec provided more heterogeneous recommendations than d2v:CF, indicating its inferior capability of precisely modeling a user. The results of the other methods (LSA, LDA, word2vec) are worse than user2vec, TF-IDF and d2v:CF.

The fact that TF-IDF and d2v:CF both model different aspects of the data and both show outstanding performance motivates to combine them. Essentially, this would follow the principle of user2vec but using separate models to encode app usage information and app descriptions. We find that constructing a hybrid filtering model through concatenating the output vectors of TF-IDF and d2v:CF outperforms both TF-IDF and d2v:CF separately significantly, as displayed in Table 8. It improves the performance of the supervised models by **2.23%** AUC-ROC (**1.26%** absolutely) compared to the zero-baseline, averaged over all models. In contrast, a concatenation of TF-IDF and user2vec is only marginally better than TF-IDF alone. These results show that modeling information on content of apps and user behaviour separately and subsequently combining models' outputs yields significantly better results than joint modeling of app information and user behaviour (user2vec). This might be due to a positive effect of keeping noisy app descriptions out of the doc2vec model training to only have it optimize on usage histories. Complementing d2v:CF, TF-IDF seems to excel at extracting key information from short and often low-quality app descriptions. However, there is more data on users and apps available than just usage histories and app descriptions that could potentially yield performance improvements.

| AUC-ROC | None | TF-IDF + d2v:CF | TF-IDF + user2vec | TF-IDF | d2v:CF | user2vec |
|---|---|---|---|---|---|---|
| male | *58.55%* | 1.18% | 1.07% | 0.95% | **1.31%** | 1.11% |
| female | *57.76%* | **1.05%** | 0.82% | 0.66% | 0.92% | 0.58% |
| 18-24 | *61.12%* | 0.54% | 0.61% | **0.69%** | 0.43% | 0.01% |
| 25-34 | *55.63%* | 1.61% | **1.63%** | 1.33% | 1.00% | 0.97% |
| 35-44 | *52.78%* | 0.70% | **0.86%** | 0.73% | 0.45% | 0.79% |
| 45-54 | *52.44%* | 0.91% | 0.86% | **0.92%** | 0.24% | 0.29% |
| 55+ | *57.25%* | **2.08%** | 1.71% | 1.79% | 1.85% | 1.30% |
| 18-34 | *56.54%* | **1.66%** | 1.40% | 1.29% | 1.17% | 0.67% |
| 35+ | *56.03%* | **1.63%** | 1.32% | 1.38% | 1.13% | 0.73% |
| Average | | **1.26%** | 1.14% | 1.08% | 0.94% | 0.72% |

Table 8: R2. Absolute change compared to the zero-baseline (None) in AUC-ROC score for combinations of TF-IDF and d2v:CF/user2vec and benchmarks.

| AUC-ROC | None | TF-IDF + context2vec | TF-IDF + d2v:CF + meta | TF-IDF + user2vec + meta | TF-IDF + d2v:CF | TF-IDF+ user2vec |
|---|---|---|---|---|---|---|
| male | *58.55%* | 0.99% | **1.19%** | 1.12% | 1.18% | 1.07% |
| female | *57.76%* | 0.98% | **1.09%** | 0.82% | 1.05% | 0.82% |
| 18-24 | *61.12%* | 0.47% | 0.62% | **0.76%** | 0.54% | 0.61% |
| 25-34 | *55.63%* | 1.66% | **1.68%** | 1.65% | 1.61% | 1.63% |
| 35-44 | *52.78%* | 0.63% | 0.69% | 0.79% | 0.70% | **0.86%** |
| 45-54 | *52.44%* | **1.02%** | 0.86% | 0.85% | 0.91% | 0.86% |
| 55+ | *57.25%* | 1.72% | 2.05% | 1.68% | **2.08%** | 1.71% |
| 18-34 | *56.54%* | 1.56% | 1.58% | 1.48% | **1.66%** | 1.40% |
| 35+ | *56.03%* | 1.46% | **1.63%** | 1.26% | 1.61% | 1.32% |
| Average | | 1.17% | **1.27%** | 1.16% | 1.26% | 1.14% |

Table 9: R3. Absolute change compared to the zero-baseline (None) in AUC-ROC score for TF-IDF + context2vec and combinations of TF-IDF and d2v:CF/user2vec excl./incl. metadata directly as features in the supervised models ("+ meta").

### 5.3 R3: Introducing Context-Awareness

Besides evaluating the performance of context2vec, we also investigate the performance change induced by directly including the metadata variables[11] in the supervised models as features. Since combinations with TF-IDF have shown to outperform all other methods, we continue to concatenate user vector representations generated by approaches that are subject to evaluation with TF-IDF vectors. In Table 9 we compare the resulting change in predictive performance between the combinations of TF-IDF with d2v:CF and user2vec excluding and including metadata as categorical/continuous numeric variables, and TF-IDF with context2vec. In this table we denote the direct inclusion of metadata as separate features in the supervised models (i.e. concatenation with user vector representations) as "+ meta". The results show that including additional metadata yields small improvements in predictive performance of the look-alike models. With a relative gain of 2.24% (1.27% absolutely), the best performing variant is the concatenation of TF-IDF and d2v:CF with metadata directly used as features in the supervised models (TF-IDF + d2v:CF + meta). However, analogous to TF-IDF + user2vec + meta, this method is only slightly better than its equivalent that does not include additional metadata. Interestingly, whereas TF-IDF + context2vec performs worse than TF-IDF + d2v:CF excluding and including metadata, it outperforms all combinations of TF-IDF with user2vec. This indicates that, first, the inclusion of context information is generally beneficial for the performance of downstream supervised models. Second, including context data in the training of the hybrid filtering model is even superior to directly including the data as features in the supervised models.

## 6 CONCLUSION

In this paper, we proposed user2vec and context2vec - novel hybrid filtering approaches for learning continuous representations of users through the application of the recent neural probabilistic language model algorithm doc2vec. What is more, we present further hybrid filtering variants that are the result of combinations of known collaborative and content-based approaches (TF-IDF + d2v:CF). Overall, our research has shown that generating embeddings through hybrid filtering using doc2vec has great potential to improve recommendation systems and predictive look-alike models. Guided by three research questions (R1, R2, R3), we assessed our approaches against a variety of baselines both qualitatively in recommendation system settings and quantitatively using supervised machine learning models for look-alike modeling that predict age and gender of users. We found that the joint modeling of usage patterns and item descriptions by user2vec performs excellently in all recommendation tasks, whereas the inclusion of metadata (context2vec) seems to slightly deteriorate the quality of recommendations again. Furthermore, we showed that the generated user embeddings represent highly important features in predictive look-alike modeling algorithms. However, at the same time we saw that separately modeling usage patterns and item descriptions and eventually combining their results leads to superior performance. A hybrid filtering combination of TF-IDF aggregated on user level (CBF) and d2v:CF (CF) performed best, resulting in a relative average increase in predictive performance (AUC-ROC) of 2.23%. Finally, we evaluated the impact of introducing context awareness by extending user2vec to context2vec through explicitly including additional user and app metadata in the doc2vec model training. Results indicated that this improves predictive performance of downstream supervised models. In fact, improvements are even larger than directly including the additional metadata as categorical or numerical features in the supervised models. This highlights the gain of providing the doc2vec model with contextual information.

In future, we intend to investigate modifications of the loss function to introduce a differentiation between context data and item descriptions. Furthermore, we aim to explore the application of novel neural probabilistic language models tailored to blocks of text, such as "skip-thought-vectors" [17], to recommendation systems.

## REFERENCES
[1] Gediminas Adomavicius and Alexander Tuzhilin. 2005. Toward the next generation of recommender systems: A survey of the state-of-the-art and possible

---
[11] OS, city, genre, price, number of ratings, average rating

[2] Oren Barkan and Noam Koenigstein. 2016. Item2vec: neural item embedding for collaborative filtering. In *Machine Learning for Signal Processing (MLSP), 2016 IEEE 26th International Workshop on*. IEEE, 1–6.
[3] Yoshua Bengio, Réjean Ducharme, Pascal Vincent, and Christian Jauvin. 2003. A neural probabilistic language model. *Journal of machine learning research* 3, Feb (2003), 1137–1155.
[4] David M Blei, Andrew Y Ng, and Michael I Jordan. 2003. Latent dirichlet allocation. *Journal of machine Learning research* 3, Jan (2003), 993–1022.
[5] Interactive Advertising Bureau. [n. d.]. IAB Display Trading Buyers Guide. https://iabuk.net/resources/handbooks/display-trading-buyers-guide. ([n. d.]). [Online; accessed 16-June-2017].
[6] Interactive Advertising Bureau. 2015. Mobile Programmatic Playbook. https://www.iab.com/wp-content/uploads/2015/05/MobileProgrammaticPlaybook.pdf. (2015). [Online; accessed 15-June-2017].
[7] Robin Burke. 2007. Hybrid web recommender systems. *The adaptive web* (2007), 377–408.
[8] Inc. comScore. 2016. The 2016 U.S. Mobile App Report. https://www.comscore.com/Insights/Presentations-and-Whitepapers/2016/The-2016-US-Mobile-App-Report. (2016). [Online; accessed 15-June-2017].
[9] Scott Deerwester, Susan T Dumais, George W Furnas, Thomas K Landauer, and Richard Harshman. 1990. Indexing by latent semantic analysis. *Journal of the American society for information science* 41, 6 (1990), 391.
[10] eMarketer. 2016. US Mobile App Install Ad Spending. http://www.emarketer.com/Chart/US-Mobile-App-Install-Ad-Spending-2014-2016-billions-change-of/-total-mobile-ad-spending/198131. (2016). [Online; accessed 15-June-2017].
[11] Diogo Gonçalves, Miguel Costa, and Francisco M Couto. 2016. A Flexible Recommendation System for Cable TV. *arXiv preprint arXiv:1609.02451* (2016).
[12] Mihajlo Grbovic, Vladan Radosavljevic, Nemanja Djuric, Narayan Bhamidipati, Jaikit Savla, Varun Bhagwan, and Doug Sharp. 2015. E-commerce in your inbox: Product recommendations at scale. In *Proceedings of the 21th ACM SIGKDD International Conference on Knowledge Discovery and Data Mining*. ACM, 1809–1818.
[13] Thomas Hofmann. 2004. Latent semantic models for collaborative filtering. *ACM Transactions on Information Systems (TOIS)* 22, 1 (2004), 89–115.
[14] Ganesh Iyer, David Soberman, and J Miguel Villas-Boas. 2005. The targeting of advertising. *Marketing Science* 24, 3 (2005), 461–476.
[15] László A Jeni, Jeffrey F Cohn, and Fernando De La Torre. 2013. Facing imbalanced data–Recommendations for the use of performance metrics. In *Affective Computing and Intelligent Interaction (ACII), 2013 Humaine Association Conference on*. IEEE, 245–251.
[16] Bhargav Kanagal, Amr Ahmed, Sandeep Pandey, Vanja Josifovski, Lluis Garcia-Pueyo, and Jeff Yuan. 2013. Focused matrix factorization for audience selection in display advertising. In *Data Engineering (ICDE), 2013 IEEE 29th International Conference on*. IEEE, 386–397.
[17] Ryan Kiros, Yukun Zhu, Ruslan R Salakhutdinov, Richard Zemel, Raquel Urtasun, Antonio Torralba, and Sanja Fidler. 2015. Skip-thought vectors. In *Advances in neural information processing systems*. 3294–3302.
[18] Ralf Krestel, Peter Fankhauser, and Wolfgang Nejdl. 2009. Latent dirichlet allocation for tag recommendation. In *Proceedings of the third ACM conference on Recommender systems*. ACM, 61–68.
[19] Quoc Le and Tomas Mikolov. 2014. Distributed representations of sentences and documents. In *Proceedings of the 31st International Conference on Machine Learning (ICML-14)*. 1188–1196.
[20] Deloitte LLP. 2016. Global Mobile Consumer Survey 2016: UK Cut. https://www.deloitte.co.uk/mobileuk/assets/pdf/Deloitte-Mobile-Consumer-2016-There-is-no-place-like-phone.pdf. (2016). [Online; accessed 15-June-2017].
[21] Harry Mak, Irena Koprinska, and Josiah Poon. 2003. Intimate: A web-based movie recommender using text categorization. In *Web Intelligence, 2003. WI 2003. Proceedings. IEEE/WIC International Conference on*. IEEE, 602–605.
[22] Ashish Mangalampalli, Adwait Ratnaparkhi, Andrew O Hatch, Abraham Bagherjeiran, Rajesh Parekh, and Vikram Pudi. 2011. A feature-pair-based associative classification approach to look-alike modeling for conversion-oriented user-targeting in tail campaigns. In *Proceedings of the 20th international conference companion on World wide web*. ACM, 85–86.
[23] Prem Melville, Raymond J Mooney, and Ramadass Nagarajan. 2002. Content-boosted collaborative filtering for improved recommendations. In *Aaai/iaai*. 187–192.
[24] Tomas Mikolov, Kai Chen, Greg Corrado, and Jeffrey Dean. 2013. Efficient estimation of word representations in vector space. *arXiv preprint arXiv:1301.3781* (2013).
[25] Tomas Mikolov, Quoc V Le, and Ilya Sutskever. 2013. Exploiting similarities among languages for machine translation. *arXiv preprint arXiv:1309.4168* (2013).
[26] Tomas Mikolov, Ilya Sutskever, Kai Chen, Greg S Corrado, and Jeff Dean. 2013. Distributed representations of words and phrases and their compositionality. In *Advances in neural information processing systems*. 3111–3119.
[27] Raymond J Mooney and Loriene Roy. 2000. Content-based book recommending using learning for text categorization. In *Proceedings of the fifth ACM conference on Digital libraries*. ACM, 195–204.
[28] Cataldo Musto, Giovanni Semeraro, Marco De Gemmis, and Pasquale Lops. 2015. Word Embedding Techniques for Content-based Recommender Systems: An Empirical Evaluation.. In *RecSys Posters*.
[29] Cataldo Musto, Giovanni Semeraro, Marco de Gemmis, and Pasquale Lops. 2016. Learning word embeddings from wikipedia for content-based recommender systems. In *European Conference on Information Retrieval*. Springer, 729–734.
[30] Makbule Gulcin Ozsoy. 2016. From word embeddings to item recommendation. *arXiv preprint arXiv:1601.01356* (2016).
[31] Owen Phelan, Kevin McCarthy, and Barry Smyth. 2009. Using twitter to recommend real-time topical news. In *Proceedings of the third ACM conference on Recommender systems*. ACM, 385–388.
[32] Van-Thuy Phi, Liu Chen, and Yu Hirate. 2016. Distributed representation based recommender systems in e-commerce. In *DEIM Forum*.
[33] Sanjay Purushotham, Yan Liu, and C-C Jay Kuo. 2012. Collaborative topic regression with social matrix factorization for recommendation systems. *arXiv preprint arXiv:1206.4684* (2012).
[34] Archana Ramesh, Ankur Teredesai, Ashish Bindra, Sreenivasulu Pokuri, and Krishna Uppala. 2013. Audience segment expansion using distributed in-database k-means clustering. In *Proceedings of the Seventh International Workshop on Data Mining for Online Advertising*. ACM, 5.
[35] Herbert Rubenstein and John B Goodenough. 1965. Contextual correlates of synonymy. *Commun. ACM* 8, 10 (1965), 627–633.
[36] Panagiotis Symeonidis, Alexandros Nanopoulos, and Yannis Manolopoulos. 2008. Tag recommendations based on tensor dimensionality reduction. In *Proceedings of the 2008 ACM conference on Recommender systems*. ACM, 43–50.
[37] Duyu Tang, Furu Wei, Nan Yang, Ming Zhou, Ting Liu, and Bing Qin. 2014. Learning Sentiment-Specific Word Embedding for Twitter Sentiment Classification.. In *ACL (1)*. 1555–1565.
[38] Oriol Vinyals, Alexander Toshev, Samy Bengio, and Dumitru Erhan. 2015. Show and tell: A neural image caption generator. In *Proceedings of the IEEE Conference on Computer Vision and Pattern Recognition*. 3156–3164.
[39] Jun Wang, Weinan Zhang, and Shuai Yuan. 2017. Display Advertising with Real-Time Bidding (RTB) and Behavioural Targeting. (2017).
[40] Kangning Wei, Jinghua Huang, and Shaohong Fu. 2007. A survey of e-commerce recommender systems. In *Service systems and service management, 2007 international conference on*. IEEE, 1–5.
[41] Guandong Xu, Yanchun Zhang, and Xun Yi. 2008. Modelling user behaviour for web recommendation using lda model. In *Web Intelligence and Intelligent Agent Technology, 2008. WI-IAT'08. IEEE/WIC/ACM International Conference on*, Vol. 3. IEEE, 529–532.
[42] Jin An Xu and Kenji Araki. 2006. A SVM-based personal recommendation system for TV programs. In *Multi-Media Modelling Conference Proceedings, 2006 12th International*. IEEE, 4–pp.
[43] Weinan Zhang, Shuai Yuan, Jun Wang, and Xuehua Shen. 2014. Real-time bidding benchmarking with iPinYou dataset. *arXiv preprint arXiv:1407.7073* (2014).